\documentclass{article}

\usepackage{spconf,amsmath,amsfonts,graphicx}
\usepackage{amssymb,textcomp,mathtools}
\usepackage{bm,upgreek,algorithm,hyperref}
\usepackage{multirow,booktabs,hhline,array}
\usepackage{cite,url,makecell,setspace, xcolor}
\usepackage{enumitem}
\pdfoutput=1

\def\thline{\noalign{\hrule height 1.0pt}}

\renewcommand{\vec}[1]{\bm{\mathrm{#1}}}

\title{Online Binaural Speech Separation of Moving Speakers With a Wavesplit Network}
\name{Cong Han, Nima Mesgarani}

\address{Department of Electrical Engineering, Columbia University, New York, NY}

\begin{document}
\ninept
\maketitle

\begin{abstract}
Binaural speech separation in real-world scenarios often involves moving speakers. Most current speech separation methods use utterance-level permutation invariant training (u-PIT) for training. In inference time, however, the order of outputs can be inconsistent over time particularly in long-form speech separation. This situation which is referred to as the speaker swap problem is even more problematic when speakers constantly move in space and therefore poses a challenge for consistent placement of speakers in output channels. Here, we describe a real-time binaural speech separation model based on a Wavesplit network to mitigate the speaker swap problem for moving speaker separation. Our model computes a speaker embedding for each speaker at each time frame from the mixed audio, aggregates embeddings using online clustering, and uses cluster centroids as speaker profiles to track each speaker throughout the long duration. Experimental results on reverberant, long-form moving multitalker speech separation show that the proposed method is less prone to speaker swap and achieves comparable performance with u-PIT based models with ground truth tracking in both separation accuracy and preserving the interaural cues.

\end{abstract}


\section{Introduction}
\label{sec:intro}
Speaker-independent speech separation is essential for augmented hearing technologies \cite{cherry1953some, hershey2016deep, kolbaek2017multitalker, chen2017cracking, han2019speaker}. Immersive augmented hearing requires binaural speech separation \cite{hadad2015theoretical,doclo2006theoretical,han2020real,tan2020sagrnn,feng2021estimation} which not only separates individual sources but also preserves spatial cues in stereo sounds so that subjects can perceive the correct location of sources in the space. However, in realistic scenarios, the speakers can move within the environments, which poses unique challenges for robust speech separation \cite{nikunen2017separation,taseska2017blind}. A traditional solution is the block-wise adaptation method which splits the signal into short blocks and applies common separation algorithms in each block under the assumption that the sources within each block are stationary \cite{koutvas2000blind,mukai2005real,zhang2007blind}. However, short blocks provide limited contextual information, and the method requires tracking speakers across
consecutive blocks \cite{mukai2003robust,Mukai2004BlindSS}. Recently, a deep learning approach that performs utterance-level source separation of moving speakers has been proposed \cite{han2021binaural}. The model utilizes longer spectral and spatial information and implicitly tracks the speakers within the utterance which significantly outperforms block-wise methods.

Despite this effort, robust tracking of moving speakers in long-form speech separation remains challenging due to the speaker swap problem where even though the overlapped sources can be well separated, the ordering of outputs may be inconsistent over time. For example, when separating speakers in a long mixture of A+B, the model outputs [A,B] between time $t_1$ and $t_2$ but outputs [B,A] between time $t_2$ and $t_3$. This phenomenon is common, and it can occur when speaker energy varies or a period of silence exists amid the mixture. Separating moving speakers is even more prone to the speaker swap problem than separating stationary speakers, especially at times when speakers move closer to each other in space. It is likely because similar spatial information results in speaker tracking ambiguity. To make the speaker order consistent in the sequence, some researchers have proposed a stitching-based algorithm that divides the long-form outputs into several overlapped segments, calculates the similarity between the overlapped regions in adjacent segments, and re-orders the segments \cite{chen2020continuous, li2021dual}. Others have designed a tracking network to track the segments \cite{zhang2021separating}. These methods are effective for non-causal systems but not suitable for causal systems which require low latency. 

An alternative solution for real-time systems is speaker-informed speech separation. In this method, the model is conditioned on a representation of each speaker which is used to track the speakers over time. The representation is usually a speaker-discriminative embedding such as i-vector \cite{dehak2010front} or d-vector \cite{variani2014deep}. There are multiple ways to acquire speaker representations. A general approach is to use a speaker ID neural network to compute speaker embedding from a voice snippet of the target speaker \cite{vzmolikova2019speakerbeam,delcroix2020improving,wang2018voicefilter,xiao2019single}. Speaker representations can also be derived from the mixture sound itself. The overlap ratio of speakers is typically low in natural conversions, hence a speaker ID network can extract one embedding at each time frame from the mixture \cite{han2021continuous} and combine all the embeddings into individual speaker clusters. The cluster centroids can then be used as representing the long-term speaker characteristics. Without the low overlap ratio assumption, Wavesplit \cite{zeghidour2021wavesplit} jointly trains a speaker stack and a separation stack where the speaker stack predicts an embedding per speaker at each time frame, aggregates them across the whole input, and uses the aggregated representation to guide the speaker stack. 

In this paper, we adopt the Wavesplit approach in the task of binaural speech separation of moving speakers. We address the speaker swap problem by training a speaker profile module that infers speaker representations from the mixture which is then used to reliably track each speaker over time and jointly perform speaker localization and separation. To ease the estimation of speaker-discriminative representations, we first train a speaker identification network to classify the identity of the source signals where the intermediate embeddings are used as the training target for the speaker profile module. Experimental results show that the proposed method can mitigate the speaker swap problem while achieving comparable performance with u-PIT models with ground truth tracking in both separation quality and preserving the spatial cues in long-form speech separation. Moreover, all modules in the system are causal and have low latency which allows real-time implementation critical in real-world hearable technologies.

\begin{figure*}[!t]
    \centering
    \includegraphics[width=2.0\columnwidth]{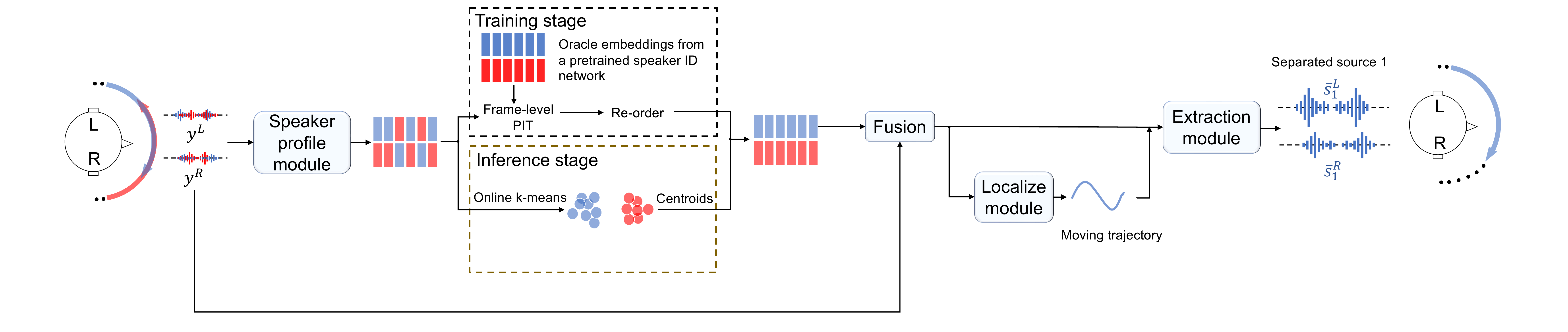}
    \caption{Schematic of the proposed system. The speaker profile module estimates an embedding for each speaker at each time frame from the binaural mixture, but the ordering of the embeddings is not necessarily consistent over time. During training, a pre-trained speaker ID network generates oracle embeddings from individual sources to serve as the training target for the speaker profile module. Frame-level permutation invariant training (PIT) is used to choose the best match and re-order the estimated embedding sequence. In inference, online k-means is performed to cluster embeddings and update the centroids. The re-ordered embedding sequence or the centroid sequence informs the localization and separation modules to jointly localize and separate the corresponding speaker. The interaural cues are preserved in the stereo output.}
    \label{fig:arch}
\end{figure*} 

% \section{Related work}
% \label{sec:rel}
% \input{relate}

\section{Method}
\label{sec:med}
\subsection{Speaker profile module}
Given the binaural mixed signals $\Vec{Y} \in \mathbb{R}^{2 \times L}$, where L is the signal length, the speaker profile module estimates N sequences of speakers vectors, $\vec{H} \in \mathbb{R}^{N \times T \times D}$, where N is the number of speakers presented in the mixture, T is the time frames, and D is the vector dimension. In this study, N is fixed to be two. $\vec{h}(n,t) \in \mathbb{R}^D $ denotes the n-th vector of $\vec{H}$ at the time frame t. It's noted that there is no predetermined ordering of the speaker embeddings at each time frame and the ordering across frames is not necessarily consistent. For example, $\vec{h}(1,t_1)$ and $\vec{h}(2,t_3)$ can represent the speaker A while $\vec{h}(2,t_1)$ and $\vec{h}(1,t_3)$ represent the speaker B as shown in Figure~\ref{fig:arch}. The embeddings are speaker-discriminative so that 1) they can be clustered into individual speaker groups; 2) each group of embeddings can guide the separation of the corresponding speaker from the mixture. 

To facilitate training, we made some modifications to Waveplit. Instead of maintaining a speaker embedding table, we trained a speaker ID network (SNet) to extract the oracle speaker embeddings from the individual sources. SNet follows the design in \cite{zhou2019cnn} and is trained 
to predict the speaker identity of [0, M-1] using the cross-entropy loss where M is the number of speaker identities in the training set. The SNet takes the source i, $\Vec{s}_i \in \mathbb{R}^{L}$, as input and outputs an embedding sequence $\vec{E_i} \in \mathbb{R}^{T \times D}$. $\vec{E} = [\vec{E}_1, ..., \vec{E}_{N}] \in \mathbb{R}^{N \times T \times D}$ is the oracle speaker embeddings. Different from $\vec{H}$, the ordering of speakers in $\vec{E}$ is consistent in the sequence. To force the embeddings to have small intra-speaker and large inter-speaker distances, we randomly sampled time frames $\{p, q\}$ and added a triplet loss, 
\begin{align}
\begin{split}
\mathcal{L}_\text{triplet } = \sum_{i,j}\sum_{p,q}\text{max}\{&|\vec{e}(i,p) - \vec{e}(i,q)| - \\&|\vec{e}(i,p) - \vec{e}(j,p)| + \text{m}, 0\},
\end{split}
\end{align}
where $\vec{e}(i,p)$ is the vector of speaker i at the time step p and m is the margin. Then, we used frame-level PIT loss to train the speaker profile module,
\begin{align}
\mathcal{L}_{\text{PIT}}(\vec{H}, \vec{E}) = \sum_{t=1}^T \min_{\pi \in P} \sum_{i=1}^N |\vec{h}(i,t) - \vec{e}(\pi(i), t)|,
\end{align}
where P is the set of all N! permutations. The best match between the oracle embeddings and estimated embeddings at each frame can be used to re-order the estimated speaker embeddings in an order consistent with the oracle ones. The re-ordered embeddings $\vec{\hat{H}}(i) \in \mathbb{R}^{T \times D}, i=1 ...N$ is the i-th speakers' profile used to guide the speech separation. 

In inference, we performed online k-means on $\vec{H}$, which keeps updating the cluster centroids over time. The sequence of the centroids $\vec{C}(i) \in \mathbb{R}^{T \times D}$ is the i-th speaker' profile.

\subsection{Speaker localization module and extraction module}
\label{sec:sem}
We trained a multi-input-multi-output (MIMO) TasNet to separate the targeted speaker. Following the design in \cite{han2021binaural}, we used 1-D convolution layers to extract the time-domain features from the waveform, and then concatenated the time-domain features and frequency-domain features, i.e., interaural phase difference (IPD) and interaural level difference (ILD) as input features which contain spectro-temporal and spatial-temporal information. The difference is that the model is conditioned by a speaker profile. Previous research has shown that feature-wise linear modulation (FiLM) is an effective conditioning method for neural networks, so we used the same method here. Specially, FiLM learns two linear functions $f_l \in \mathbb{R}^{D \times D}$ and $g_l \in \mathbb{R}^{D \times D}$ at layer $l$ which project the speaker profile $\vec{\hat{H}}(i)$ to $\gamma_{l}(i) \in \mathbb{R}^{T \times D}$ and $\beta_{l}(i) \in \mathbb{R}^{T \times D}$, respectively,
\begin{align}
\gamma_{l}(i) = \vec{\hat{H}}(i)\cdot f_l, \quad
\beta_{l}(i) = \vec{\hat{H}}(i) \cdot g_l,
\end{align}
where $\gamma_{l}(i)$ and $\beta_{l}(i)$ modulate the activation $x_l$ at layer i,
\begin{align}
\text{FiLM}(x_l|\gamma_{l}(i), \beta_{l}(i)) = \gamma_{l}(i) \times x_l + \beta_{l}(i).
\end{align}
We added one FiLM before each convolutional block in TasNet.

The system jointly localizes and separates the target speaker. The localization module performs the classification of the direction of arrival (DOA) at each time frame. We discretized the DOA angles into K classes. The localization module estimates a classification matrix $\vec{V}(i) \in \mathbb{R}^{T \times K}$ for the speaker i. To train the localization module, we computed the cross-entropy loss between $\vec{V}(i)$ and the target DOA labels. 
We split $\vec{V}(i)$ into B small chunks with each chunk containing Q time frames, $T = B \times Q$, and count the frequency of each DOA class in each chunk, and consider the most frequent class as the estimated DOA for that chunk. The explicitly estimated trajectories enable us to determine the moving source we are interested in and to modify the acoustic scene accordingly, for example, by amplifying or attenuating individual sources. 

We concatenated $\vec{V}(i)$ with other fusion features to extract the target speaker $\hat{\vec{S}}_i = [\hat{\vec{s}}^L, \hat{\vec{s}}^R]$, where $\hat{\vec{s}}_i^L$ and $\hat{\vec{s}}_i^R$ are the estimated left- and right-channel signals of the source i. Since the target speaker is determined by $\vec{\hat{H}}(i)$, there is no permutation problem. The reconstruction loss $\mathcal{L}_\text{extraction}$ is:
\begin{align}
\label{eqn:snr}
    &\mathcal{L}_\text{extraction} = \text{SNR}(\vec{s}_i^L, \hat{\vec{s}}_i^L) + \text{SNR}(\vec{s}_i^R, \hat{\vec{s}}_i^R), \\
    &\text{SNR}(\vec{x}, \hat{\vec{x}}) = 10\,\text{log}_{10}\left(\frac{||\vec{x}||_2^2}{||\hat{\vec{x}} - \vec{x}||_2^2}\right).
\end{align} 
Because SNR is sensitive to both the time shift and power scale of the estimated waveform, it can force the ITD and IPD to be preserved in the estimated waveform  \cite{han2020real}. 

The speaker profile module, localization module, and extraction module were trained separately and then trained jointly.

\section{Experimental settings}
\label{sec:exp}
\subsection{Dataset}
\subsubsection{Binaural room impulse responses and speech data}
We used two types of binaural room impulse responses (BRIRs) in this study: one obtained from simulated rooms and the other one measured in real rooms, \footnote{\url{http://iosr.uk/software/index.php}}. There are 11 simulated rooms with reverberation time (RT60) varying from 0 to 1 s with 0.1 s increments. In this study, we only chose 8 rooms with RT60 from 0 to 0.7 s. There are 4 real rooms with RT60 0.32 s, 0.47 s, 0.68 s, 0.89 s, respectively, and we chose the first three. The impulse responses were calculated with the sound source located on the frontal azimuthal plane between $-90^{\circ}$ and $90^{\circ}$ with $5^{\circ}$ increments at a distance of 1.5 m to the receiver. Two speakers were randomly selected from the 100-hour Librispeech dataset \cite{panayotov2015librispeech}. Both speech data and BRIRs were sampled at 16 kHz.

\subsubsection{Moving source simulation}
Given a monaural speech $\vec{s}$ and a set of BRIRs $\left\{h_j^L\right\}_{j=1}^N, \left\{h_j^R\right\}_{j=1}^N \in \mathbb{R}^{L_h}$, where $h_j^L$ and $h_j^R$ are the BRIR filters from the location j to the left ear and right ear, respectively, $L_h$ is the filter length, and N is the number of locations (37 in this study), the moving binaural source was simulated as:
\begin{align}
    &\vec{s}^L[n] = \sum_{j=0}^{N}\sum_{k=0}^{L_h} \mathbb{I}_j(n)\cdot h_j^L[k] \cdot s[n-k] \\
    &\vec{s}^R[n] = \sum_{j=0}^{N}\sum_{k=0}^{L_h} \mathbb{I}_j(n) \cdot h_j^R[k] \cdot s[n-k]
\end{align}
where $\mathbb{I}_j(n)$ is an indicator function. $\mathbb{I}_j(n) = 1$ when and only when $\vec{s}$ is at location j at time step n, $\mathbb{I}_j(n) = 0$ in other cases.  %In this study, we omit the doppler effects \cite{smith2002doppler} and HRTF interpolation for simplicity.

We simulated 26.7 hours and 6.7 hours of data for the training and validation, respectively using only the simulated BRIRs. For testing on simulated rooms, we created 4 hours of data using the simulated BRIRs. For testing on real rooms, we generated 1 hour of data for each of the three rooms with RT60 0.32 s, 0.47 s, 0.68 s, respectively. In this study, each mixture contained two moving speakers. The speech signals were rescaled to a random relative SNR between 0 and 5 dB. The moving velocity of each speaker was randomly chosen between 8 and $15^{\circ}/s$ and moving direction was randomly chosen between clockwise and counter-clockwise.

\begin{figure}[!t]
    \centering
    \includegraphics[width=1.0\columnwidth]{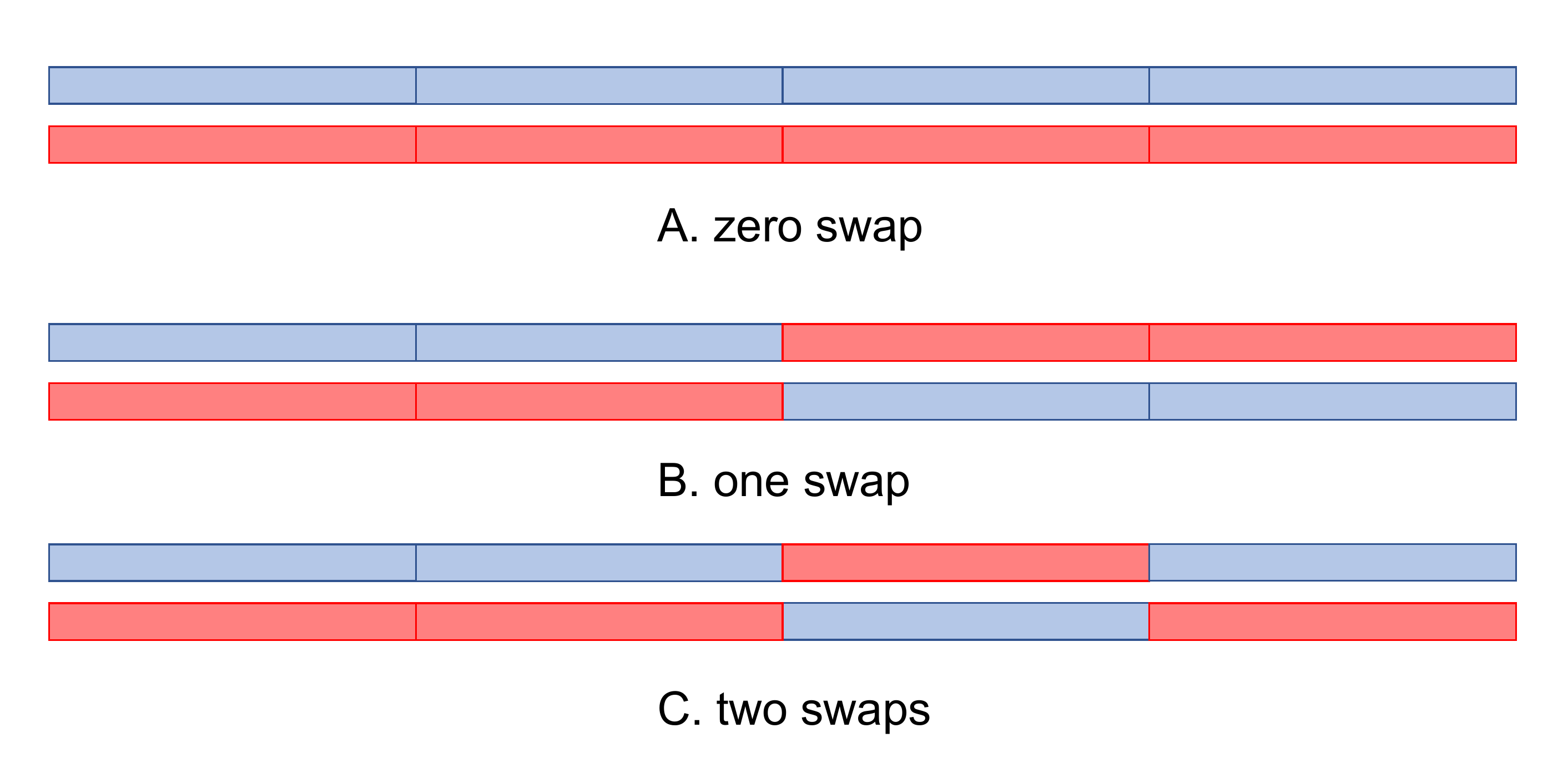}
    \caption{Examples of speaker swap in separation outputs. Blue bar denotes the speaker 1 and red bar denotes the speaker 2.}
    \label{fig:f2}
\end{figure}

\subsection{Evaluation metrics}
We evaluated the models by measuring the separation quality, the preservation of spatial cues, and the robustness to speaker swap problem. We used SNR as the signal quality metric. Following \cite{han2021binaural}, we trained a speaker localizer to examine the locational information encoded in stereo outputs. The localizer predicts the DOA every 80 ms. We calculate absolute DOA errors as the metric for the accuracy of preserving spatial cues. It is noted the localization model for evaluation here is different from the one in Section~\ref{sec:sem} as the latter one aims to decode the trajectory of the target speaker to facilitate the separation. When evaluating long-form speech separation, we divide the separation outputs into N segments and check if the order of outputs in adjacent segments is consistent. We count the number of speaker swaps in separation outputs as the metric for robustness to the speaker swap problem. Figure~\ref{fig:f2} shows three examples. The duration of long recordings is 24 seconds, and N is 10.

\subsection{Network architectures}
All the modules except the localization module are based on the causal configuration of TasNet \cite{luo2019conv}. We used 64 filters in the linear encoder and decoder with 4 ms filter length (i.e. 64 samples at 16 k Hz). We used 5 repeated stacks for the speaker profile module and speaker ID network, 2 repeated stacks for the fusion module, and 3 repeated stack for the extraction module with each stack having seven 1-D convolutional blocks. The localization module is a two-layer uni-directional LSTM.

\subsection{Models for comparison}
We used several monaural and binaural separation models for comparison. The single-input-multi-output (SIMO) TasNet is a monaural model that separates the mixture in left- and right-channels independently. Block-wise MIMO TasNet and uPIT-MIMO TasNet \cite{han2021binaural} are binaural baselines. Block-wise MIMO TasNet separates speech in each short block independently and concatenates the block outputs using oracle tracking information. SPK-MIMO TasNet is the proposed method in this paper, and oracle SPK-MIMO TasNet means the model is conditioned on the oracle speaker profiles from the speaker ID network. We add the same post binaural speech enhancement \cite{han2021binaural} after each binaural separation model for comparison, denoted as ``+enh" in Table~\ref{tab:t1}. When evaluating models on long recordings, we divided the separation outputs of uPIT-MIMO TasNe into segments as shown in Figure~\ref{fig:f2} and re-ordered the segments using the ground truth signals, which is referred to as uPIT-MIMO TasNet w/ tracking in Table~\ref{tab:t2} and Table~\ref{tab:t3}.

\section{Results and discussions}
\label{sec:results}
\begin{table}[!ht]
    \small
	\centering
	\caption{Experimental results of moving source separation on 2.4-second recordings.} 
    \begin{tabular}{c|c|c|c}
    \thline
    Method & \thead{Context \\ size (s)} & \thead{SNR \\ (dB) } & \thead{DOA \\ error (\textdegree)}\\
    \thline
    \multicolumn{1}{l|}{Unprocessed} & - & 0 & -\\
    \thline
    \multicolumn{1}{l|}{SIMO TasNet} & 2.4 & 5.1 & 20.9 \\
    \hline
    \multicolumn{1}{l|}{\multirow{3}{*}{Block-wise MIMO TasNet}}
	&   0.1 & 5.7 & 16.3 \\
	& 0.2 & 6.0 & 15.4 \\
	& 0.3 & 6.2 & 14.1 \\
	\hline
	\multicolumn{1}{l|}{uPIT-MIMO TasNet} & \multirow{3}{*}{2.4} & 8.4 & 8.3\\
	\multicolumn{1}{l|}{\bf{SPK-MIMO TasNet}} 
	& & 8.3 & 8.2 \\
	\multicolumn{1}{l|}{Oracle SPK-MIMO TasNet} 
	& & 8.9 & 7.4 \\
	\thline
	\multicolumn{1}{l|}{uPIT-MIMO TasNet + Enh} & \multirow{3}{*}{2.4} & 9.4 & 6.1\\
	\multicolumn{1}{l|}{SPK-MIMO TasNet + Enh} & & 9.4 & 6.0 \\
	\multicolumn{1}{l|}{Oracle SPK-MIMO TasNet + Enh} & & 9.6 & 5.8 \\
	\thline
    \multicolumn{1}{l|}{Reverberant clean} & - & & 0.5 \\
    \thline
    \end{tabular}
    \label{tab:t1}
\end{table}

Table~\ref{tab:t1} compares different methods for moving source separation on 2.4s recordings. uPIT-MIMO TasNet outperforms both the SIMO TasNet and block-wise adaptation methods by a large margin because uPIT-MIMO TasNet takes advantage of longer spectral-temporal and spatial-temporal information for moving speaker separation. Oracle SPK-MIMO TasNet, conditioned on the oracle speaker profile, achieves 0.5 dB SNR gain over uPIT-MIMO TasNet, proving the effectiveness of speaker-informed speech separation in moving speaker cases. The performance drops slightly when we use the speaker profiles inferred from the mixture. It is different from \cite{zeghidour2021wavesplit} where Wavelplit using inferred speaker representation greatly outperforms uPIT based models. One explanation is that Wavesplit employs various forms of regularization to improve the generalization capability during training. The other is that inferring long-form speaker representations that are only related to voice characteristics from moving speakers is more challenging, especially in reverberant environments. Moreover, our system is in causal configuration, and the speaker profiles are updated over time, so the speaker profiles become stable after several time frames. In real applications, we can select the source of interest based on the speaker representation, decoded moving trajectory, and the separated waveforms and employ a post binaural speech enhancement module to enhance the result. We notice that a post binaural speech enhancement module lets both uPIT-MIMO TasNet and SPK-MIMO TasNet improve the SNR and reduce the DOA error. 

\begin{table}[!h]
    \small
	\centering
	\caption{Experimental results on 24-second long recordings.}
    \begin{tabular}{c|c|c|c}
    \thline
    Method & \thead{\# of \\swaps} & \thead{SNR \\ (dB) } & \thead{DOA \\ error (\textdegree)} \\
    \thline
    uPIT-MIMO TasNet w/o tracking & 3.4 & 1.2 & 35.3 \\
    uPIT-MIMO TasNet w/ \,\, tracking & 3.4 & 7.6 & 10.0 \\
    \bf{SPK-MIMO TasNet} & 0.6 & 7.7 & 9.3\\
    oracle SPK-MIMO TasNet & 0.4 & 8.2 & 8.1\\
    \thline
    \end{tabular}
    \label{tab:t2}
\end{table}

Table~\ref{tab:t2} compares methods on 24 s recordings. We see that uPIT-MIMO TasNet has multiple times of speaker swaps in long-form speech separation, which severely affects the overall SNR. The speakers are separated but not consistently placed in certain output channels. With the ground truth tracking, the overall SNR is improved from 1.2 dB to 7.5 dB and the overall DOA error is reduced from 35.3\textdegree to 10.0\textdegree. We see that SPK-MIMO TasNet and oracle SPK-MIMO TasNet, which are conditioned on a speaker representation, are less prone to speaker swap and tends to assign each speaker into a certain output channel. The model conditioned on the inferred speaker profiles is slightly worse than that conditioned on the oracle speaker profile but is slightly better than u-PIT based model with tracking. This shows the proposed method is more robust with respect to a targeted moving signal.

\begin{table}[!t]
    \scriptsize
	\caption{Experimental results on 24-second long recordings in different rooms. The number of swaps / SNR (dB) / DOA error (\textdegree) are reported.}
	\begin{tabular}{c|c|c|c}
		\thline
		\multirow{2}{*}{\thead{Model}} & \multicolumn{3}{c}{\thead{Rooms}} \\
		\cline{2-4}
		& A (0.32 s) & B (0.47 s) & C (0.67 s) \\
        \thline
        uPIT-MIMO TasNet w/o tracking & 3.6/1.3/37.2 & 3.5/1.2/36.4 & 3.6/1.0/38.0 \\
        uPIT-MIMO TasNet w/ \,\, tracking & 3.6/7.0/11.6 & 3.5/6.1/15.1 & 3.6/6.0/16.3 \\
        \bf{SPK-MIMO TasNet} & 0.6/7.2/10.0 & 0.8/6.3/13.4 & 0.6/6.2/13.2\\
        oracle SPK-MIMO TasNet & 0.3/7.6/9.6 & 0.4/6.5/12.1 & 0.4/6.4/12.9 \\
        \thline

    \end{tabular}
    \label{tab:t3}
\end{table}

We also compared the models on 24 s recordings simulated using real room impulse responses with different reverberant time as shown in Table~\ref{tab:t3}. All the model are trained in the simulated rooms but can generalize to the real rooms well. Stronger room reverberation deteriorates the performance of the model in terms of the signal quality and accuracy of the preserved spatial cues. However, room reverberation has little impact on the number of speaker swaps.

\section{Conclusion}
\label{sec:con}
This paper investigates the problem of binaural speech separation of moving speakers with preserving interaural cues for long recordings. We adopted and modified the Wavesplit approach to address the speaker swap problem. 
We trained a speaker profile module to estimate an embedding per speaker at each time frame from the mixture, and used online k-means to aggregate embeddings into individual speaker profiles. Then, the system is conditioned on each speaker profile to localize and separate the speaker jointly. Objective evaluation experiments show that the proposed method mitigates the swap problem while achieving on par performance with uPIT-based models with ground truth tracking. Furthermore, the model trained with simulated room impulse responses could successfully generalize to real rooms acoustics. Future works include handling unknown and changing number of overlapped speakers and developing an efficient and compact implementation of the model that can fit into various hearable devices which require low power and low storage models.

%These makes our method suitable for real-world speech separation scenarios in which talkers are freely moving in space. %

\section{Acknowledgements}

{This work was funded by the national institute of health (NIH-NIDCD) and a grant from Marie-Josée and Henry R. Kravis.}

\newpage
\bibliographystyle{IEEEtran}

{\footnotesize\bibliography{mybib}}

\end{document}